\documentclass{aa}
\usepackage{graphics}
\usepackage{times}

\newcommand{\ltsima} {$\; \buildrel < \over \sim \;$}
\newcommand{\simlt}  {\lower.5ex\hbox{\ltsima}}            
\newcommand{\gtsima} {$\; \buildrel > \over \sim \;$}
\newcommand{\simgt}  {\lower.5ex\hbox{\gtsima}}            

\begin{document}

\thesaurus{
08(13.18.5; 
08.09.2 \object{Cygnus~X-3}; 
08.09.2 \object{Cygnus~X-1}; 
08.09.2 \object{GRS~1915+105}; 
08.09.2 \object{LSI+61$^{\circ}$303};
08.09.2 \object{SS~433};
)}

\title{Exploring the high frequency emission of radio loud X-ray binaries}

\author{J.M. Paredes\inst{1}
\and J. Mart\'{\i}\inst{2}
\and M. Peracaula\inst{3,1}
\and G. Pooley\inst{4}
\and I.F. Mirabel\inst{5,6}
}

\institute{Departament d'Astronomia i Meteorologia, Universitat de
Barcelona, Av.  Diagonal 647, E-08028 Barcelona, Spain
\and
Departamento de F\'{\i}sica, Escuela Polit\'ecnica
Superior, Universidad de Ja\'en, Calle Virgen de la Cabeza 2, E-23071 Ja\'en, Spain
\and
Department of Physics and Astronomy, University of Calgary, Calgary, Alberta, T2N IN4, Canada
\and
Mullard Radio Astronomy Observatory, Cavendish Laboratory, Cambridge, U.K.
\and
CEA/DSM/DAPNIA/Service d'Astrophysique, Centre d'\'Etudes de Saclay,
F-91191 Gif-Sur-Yvette, France
\and
Instituto de Astronom\'{\i}a y F\'{\i}sica del Espacio, C.C. 67, Suc. 28, 1428 Buenos Aires, Argentina
}

\offprints{J.M. Paredes, josepmp@mizar.am.ub.es}

\date{Received    / Accepted                  }

\maketitle

\begin{abstract}
We report millimetre-wave continuum observations of the X-ray binaries
Cygnus~X-3, SS~433, LSI+61$^{\circ}$303, Cygnus~X-1 and GRS~1915+105.
The observations were carried out with the IRAM 30 m-antenna at
250 GHz (1.25 mm) from 1998 March 14 to March 20. 
These millimetre measurements are 
complemented with centimetre observations from the Ryle Telescope, at 15 GHz (2.0 cm), and from 
the Green Bank Interferometer at 2.25 and 8.3 GHz (13 and 3.6 cm). 
Both Cygnus X-3 and SS 433 underwent moderate flaring events during our
observations, whose main spectral evolution properties are described
and interpreted. A significant spectral steepening was observed in both sources
during the flare decay, that is likely to be caused by adiabatic expansion, inverse
Compton and synchrotron losses.
Finally, we also report 250 GHz upper limits for three
additional undetected X-ray binary stars: 
\object{LSI+65$^{\circ}$010}, \object{LSI+61$^{\circ}$235} and \object{X Per}.


\end{abstract}


\section{Introduction} \label{intro}

An interesting group of X-ray binary systems in our Galaxy are known to be 
powerful and efficient sources of radio waves.
The number of radio emitting X-ray binaries (REXRBs) detected so 
far is about 10 $\%$ of the total $\sim$ 200 systems catalogued (Hjellming \&
Han 1995). Although far from representing a numerous population, their remarkable properties and
scaled down similarity with extragalactic AGNs and quasars makes them to deserve a careful
study based on multi-wavelength monitoring programs.

Radio emission in REXRBs is normally highly variable and of non-thermal synchrotron origin.
Radio outbursts with different amplitude are frequently detected and interpreted
as synchrotron radiation due to the ejection and expansion of ionized plasma clouds (plasmons), usually
following a super Eddington accretion event. 
Recent multi-wavelength monitoring of radio outbursts from the microquasar system
GRS 1915+105 (Fender et al. 1997a; Mirabel et al. 1998) have revealed that the flaring synchrotron emission
extends well beyond the centimetric domain, reaching up to infrared
wavelengths. The energetic implications of this fact are considerable (Mirabel et al. 1998),
and it would be important to investigate if similar behaviour is observed in other REXRBs. 
In an attempt to better assess this issue, we undertook 
a daily monitoring campaign in the mm domain for some well known objects in the REXRB class. 
Our main goal here was to study the variability and spectral index
properties of selected sources in a wide frequency range.
Whenever possible, we have taken advantatge from the 
availability of daily monitorings in the cm domain, 
thanks to the Ryle Telescope and to the Green Bank Interferometer.   
This allowed us to estimate the source spectral indices between radio frequencies
separated by two orders of magnitude.




The targets for the observing program were chosen among the brightest REXRBs 
with luminous massive companions and declination $\delta > -30^{\circ}$. They include:
Cygnus X-3, SS 433, LSI+61$^{\circ}$303, Cygnus X-1 and GRS~1915+105.
A summary of their main physical properties is condensed in Table~\ref{properties}. 
Previous cm observations for all of them are abundant in the literature, and a few
mm detections have been reported as well.  Nevertheless, 
no extended and truly simultaneous cm/mm monitoring
has been systematically carried out to our knowledge. The present
work is an exploratory step in this direction.

\begin{table*}
\caption[]{\label{properties} REXRBs properties}
\begin{tabular}{lccccc}
\hline
Source          & Classification       &  cm-Flux density (Jy)   & mm-Flux density (Jy) & Periods    &  Remarks   \\
                &  & Quiescent~~~Outburst & Quiescent~~~Outburst     &           \\
\hline

Cygnus X-3  &   W-R+? &  $\sim$0.05~~~$>$20  & $\sim$0.05~~~$>$3 
            &  4.8 h (orbital)  &    jets  \\
Cygnus X-1  &   O9.7Iab+BH & $\sim$0.015~~~$\sim$0.040 & 0.010
            &   5.6 d (orbital)               &            \\ 
GRS~1915+105&  Be?+BH & $\sim$0.01~~~~$>$1 & 0.015
            &                   &  superluminal jets        \\
LSI+61$^{\circ}$303& Be+NS?   &$\sim$0.03~~~~$>$0.5 & $\sim$0.01 &26.5 d (orb.), 4 yr ?                   &  radio and X-ray period        \\
SS~433      &   OB?+?      &   $\sim$0.5~~~~$>$10          &  $\sim$0.12
            &  13 d (orb.), 164 d                 &   precessing jets       \\
\hline

\end{tabular}
~\\
\end{table*}

\begin{table}
\caption[]{\label{obsres} IRAM results at 250 GHz (1.25 mm)}
\begin{tabular}{lccc}
\hline
Source          & Date         &  Number  of   & Flux density    \\
                & (1998 March) &  subscans     &     (mJy)       \\
\hline

Cygnus X-3  &   14.532 &        60   &        $44\pm 4$    \\
            &   14.555 &        19   &        $47\pm 9$    \\
            &   15.531 &       124   &        $28\pm 4$    \\
            &   16.555 &        84   &        $32\pm 4$    \\
            &   18.569 &        60   &        $91\pm 6$    \\
            &   19.571 &        93   &        $31\pm 5$    \\
            &   20.558 &        60   &        $45\pm 7$    \\

\\

Cygnus X-1  &   14.530  &        120   &      $<$  9        \\
            &   15.555  &        100   &      $<$ 17        \\
            &   16.537  &        100   &      $<$ 12        \\
            &   18.583  &        100   &      $<$ 18        \\
            &   19.587  &         50   &      $<$ 24        \\
            &   20.575  &        102   &      $<$ 21        \\

\\

GRS 1915+105  & 15.510  &         29   &      $<$ 27        \\
              & 16.508  &        126   &      $<$ 12        \\
              & 18.543  &        100   &      $<$ 37        \\
              & 19.543  &         32   &      $<$ 47        \\
              & 20.519  &         61   &      $<$ 23        \\

\\

LSI+61$^{\circ}$303   &   14.500  &         60   &      $<$ 12        \\
                      &   14.715  &        224   &      $9\pm3$       \\
                      &   15.591  &        140   &      $14\pm4$      \\
                      &   16.588  &        107   &      $<$ 12        \\
                      &   17.705  &        300   &      $<$  6        \\
                      &   18.610  &        100   &      $<$ 15        \\
                      &   19.612  &         90   &      $<$ 12        \\
                      &   20.610  &        160   &      $<$ 11        \\
\\

SS 433      &   15.497  &         24   &        $70\pm 7$   \\
            &   16.491  &         50   &        $68\pm 7$   \\
            &   18.525  &        100   &        $51\pm10$   \\
            &   19.525  &        100   &        $48\pm 6$   \\
            &   20.508  &         28   &        $72\pm10$   \\
\hline

\end{tabular}
~\\
\end{table}

\section{Observations} \label{sdos}

The observations of the REXRBs listed in Table \ref{properties} were carried out,
during the interval 1998 March 14 to March 20, using the
following astronomical facilities:

\subsection{IRAM 30 m-telescope}

Observing sessions were conducted at the 30 m telescope 
of the Institut de Radio Astronomie Millim\'etrique (IRAM) in Pico Veleta (Spain).
The backend installed was the Max Planck Institut f\"ur Radioastronomie 
(MPIfR) 39 channel bolometer array,
operating at the 250 GHz frequency (1.25 mm). The technical details of this instrument
and those of the 30 m telescope are described by Wild (1995) and Kramer et al. (1998). 
Each daily session was started with a skydip, pointing and focus sequence. Such sequence
was repeated every hour minimum and very often before a new target source. In this way,
we were able to ensure an continuous monitoring on the atmospheric opacity, pointing offsets and
focusing parameters of the telescope.
The method used to determine the pointing corrections was by cross-scanning on a
bright calibrator as close as possible to the target source.
For the program sources, the ON-OFF technique was preferred in order to obtain higher sensitivity.
The ON-OFF procedure should also help to remove possible background extended emission
around some of our sources, specially in the case of SS 433 and its associated radio nebula W50. 
A typical target observation consisted of several tens of symmetric ON-OFFs subscans,
during which the antenna beam was nutated on and off the source by 46$\arcsec$  
at a sampling rate of 0.5 s. The duration of each subscan was 10 s.
The resulting count rate, for a given source, is taken as the weighted average
of all these individual subscans.
Source counts were finally converted to flux density using planets as calibrators. 

The data reduction was all carried out with the   
New Imaging Concept (NIC) package, available at the telescope site
and originally written at the MPIfR and IRAM. 
The NIC tasks include spike removal, bad subscan flagging, 
gain channel and elevation corrections, and correlated channel noise subtraction.

\subsection{Ryle Telescope}

The Ryle Telescope at the Mullard Radio Astronomy Observatory (MRAO) was used to monitor the daily 
flux variations of four sources in our list: Cygnus X-3, Cygnus X-1,  GRS~1915+105 and LSI+61$^{\circ}$303. 
The Ryle Telescope operates at the frequency of 15 GHz (2.0 cm). Details of the
observing procedure are given in Pooley \& Fender (1997). The data reported here
were measured with linearly-polarized feed-horns and represent the Stokes parameters $I+Q$. 

\subsection{Green Bank Interferometer}

The public data from the Green Bank Interferometer\footnote{
The GBI is a facility of the USA National Science Foundation operated by the NRAO
in support of NASA High Energy Astrophysics programs.}
(GBI) were also retrieved to complete our study.
The GBI consists of two 26 m antennas on a 2.4 km baseline observing 
simultaneously at 2.25 and 8.3 GHz (13 and 3.6 cm).
The data taken are made available to the public immediately. 
All our targets are included among the radio sources routinely
monitored by the GBI. Some of them are also observed more than five times daily. 
Typical errors in GBI data are 4 mJy at
2.25 GHz and 6 mJy at 8.3 GHz for fluxes less than 100 mJy. For fluxes of about
1 Jy (as in SS433), the errors at 2.25 and 8.3 GHz are 15 and 50 mJy respectively. In order to
better compare the GBI light curves with those obtained
at other instruments, the GBI observations were averaged on a daily basis.

\begin{figure*}[htb]
\mbox{}
\vspace{6.4cm}
\includegraphics{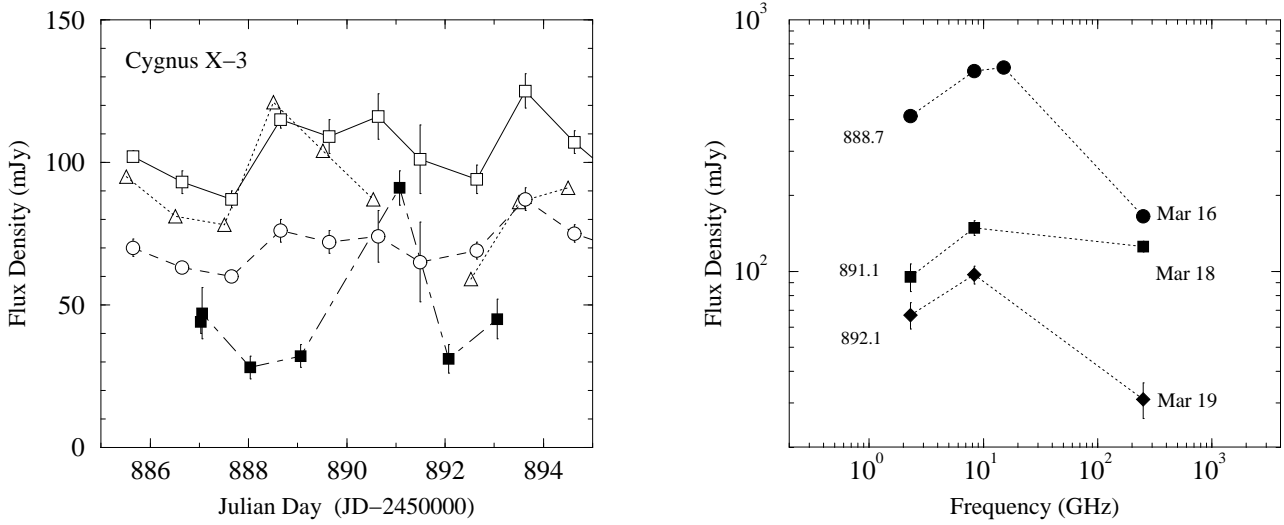}
\caption[]{ {\bf Left.} Millimetre and centimetre light curves of Cygnus X-3
as observed by IRAM 30 m antenna (250 GHz, \rule{2mm}{2mm}), GBI (2.3
GHz, $\bigcirc$, and 8.3 GHz, $\sqcap$\hspace{-2.1mm}$\sqcup$) and
Ryle Telescope (15 GHz, $\bigtriangleup$).
Error bars not shown are smaller than the symbol size.
{\bf Right.} Spectral evolution of Cygnus X-3 for some selected epochs.
An arbitrary vertical offset has been added to the first and second spectra
for better clarity display.}
\label{cx3}
\end{figure*}

\section{Results}

The summary of our main results, those obtained with the IRAM 30 m-telescope at 250 GHz (1.25 mm),
is presented in Table~\ref{obsres}. First column lists the star name; the second one indicates
the date of the observation and the third column lists the number of subscans performed.
Finally, fourth column gives the measured flux density
and its error.  The error quoted is based on the standard deviation for the
estimated flux density. For the cases where the radio star has not been
detected, an upper limit of three times the rms noise is given.
Among the sources listed in Table~\ref{properties},
the only ones that we could never detect at 250 GHz were Cygnus X-1 and GRS 1915+105.
In the cm domain, all targets were found to be detectable with the Ryle and GBI facilities.
The cm flux density levels appeared mostly consistent with
expectations based on the available literature. 

\begin{figure*}[htb]
\mbox{}
\vspace{8.4cm}
\includegraphics{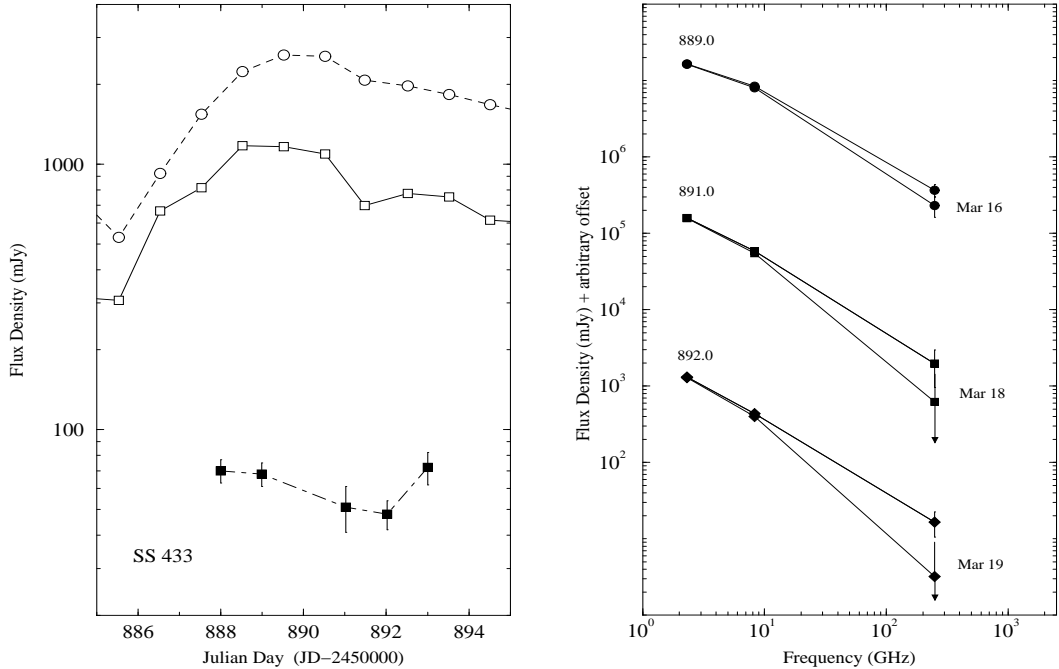}
\caption[]{
{\bf Left.} Observed millimetre and centimetre light curves of SS 433 in logarithmic scale and
with the same notation as in Fig. 1. The development of a radio outburst event is clearly seen.
{\bf Right.} Spectral evolution of the source for some selected epochs. In this right panel,
the persistent quiescent emission of SS 433 has been subtracted as estimated from both Seaquist et al. (1982)
and contemporaneous GBI data. The subtraction procedure
provides then two estimates of the SS 433 flaring emission for each epoch, i.e., lower and upper, respectively.
}
\label{ss433}
\end{figure*}

\begin{figure*}[htb]
\mbox{}
\vspace{6.4cm}
\includegraphics{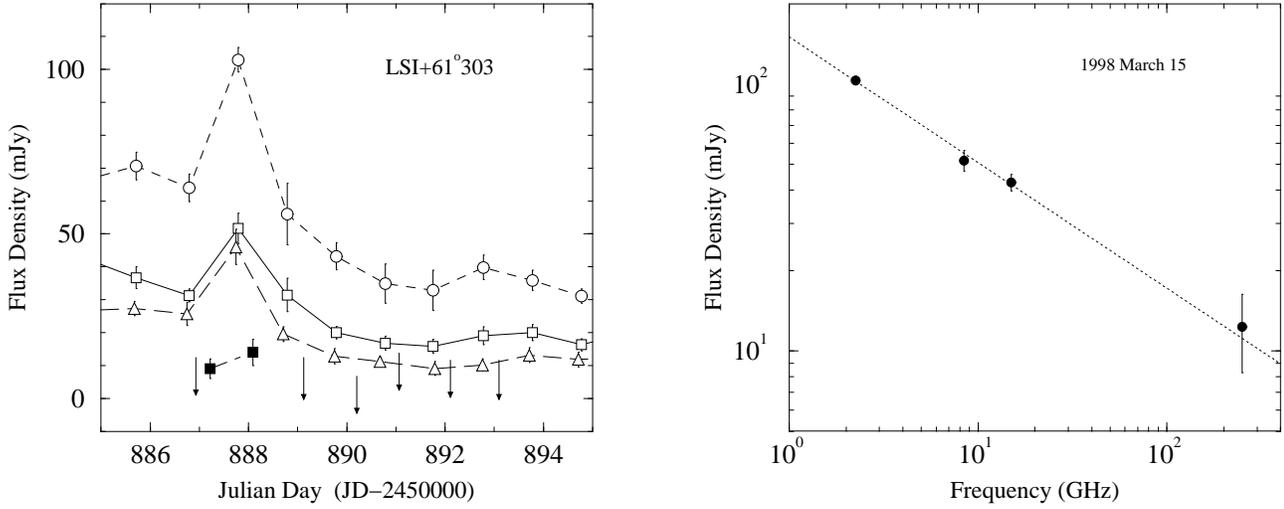}
\caption[]{
{\bf Left.} Millimetre and centimetre light curves of LSI+61$^{\circ}$303
with the same notation as in the previous Figure. The arrows indicate the
corresponding upper limits of three times the rms noise when no detection was achieved.
{\bf Right.} Radio spectrum of LSI+61$^{\circ}$303 around the maximum of
its radio outburst, extending from cm to mm wavelengths. The dashed line
is a power law fit resulting in a spectral index $\alpha=-0.47\pm0.01$.}
\label{lsi}
\end{figure*}

The light curves for Cygnus X-3, SS 433 and LSI+61$^{\circ}$303  
provided several positive mm detections and we additionally plot them in 
Figs. \ref{cx3}, \ref{ss433} and \ref{lsi}. These figures also include panels with
selected radio spectra from cm to mm wavelengths. They have
been computed by interpolating (when possible) the daily Ryle and GBI monitorings at the times  
of the corresponding IRAM observations. When discussing this spectral information, 
the spectral index $\alpha$ will be defined as $S_{\nu} \propto \nu^{\alpha}$, 
where $S_{\nu}$ is the flux density and $\nu$ the frequency. 


\subsection{Cygnus X-3}

\begin{figure}[htb]
\mbox{}
\vspace{7.5cm}
\includegraphics{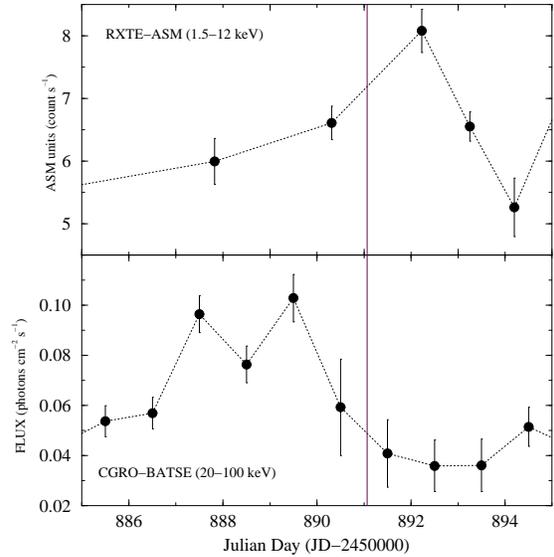}
\caption[]{The HXR and SXR emission of Cygnus X-3 during
the dates of our radio observations, as retrieved from the RXTE-ASM
and CGRO-BATSE public monitorings. The vertical line indicates
the epoch when the highest mm flux density was observed, that
we interpret as a low level radio flare.}

\label{highe}
\end{figure}

This system is currently regarded as a Wolf-Rayet star plus a compact object 
(van Kerkwijk et al. 1992; Fender et al. 1999a). 
The orbital cycle is assumed to be 4.8 h
based on the strong modulation observed with this period, specially in the X-ray domain
(Parsignault et al. 1972).
This modulation has been also reported at infrared (Becklin et al. 1973) and
radio wavelengths (Molnar 1985).
Cygnus X-3 is well known for its strong radio flares
reaching cm and mm peak flux densities of several Jy (Gregory et al. 1972 and 
references therein; Nesterov 1992), that may be interpreted
in terms of collimated ejection events (e.g. Mart\'{\i} et al. 1992). Waltman et al. (1994) 
have shown that Cygnus X-3 exhibits periods of normal quiescent emission, varying on time scales of
months from 60 to 140 mJy. The average emission level is often 
around 80 mJy and 90 mJy at 2.25 GHz and 8.3 GHz, respectively.
From the GBI data presented in Fig.~\ref{cx3}, it is clear that 
Cygnus X-3 was in a quiescent state during our observations 
with only minor flaring events taking place.

At mm wavelengths, 
only five observations have been published in the past. The first two were carried out
during a strong flaring period in the radio (Pompherey \& Epstein 1972; Baars et al. 1986).
Fender et al. (1995) and Tsutsumi et al. (1996) do detect the Cygnus X-3 quiescent radio emission 
at mm wavelengths using the JCMT. A low level
flare peaking at 263 mJy was detected by Altenhoff et al. (1994) with the IRAM
30 m-telescope at 250 GHz. Finally, there is also a sub-mm detection of Cygnus X-3
by Fender et al. (1997b). 

The data in this paper, that extend over nearly one week, 
confirm that Cygnus X-3 is continuously active 
at 250 GHz even during quiescent emission periods. The peak frequency of the spectrum
is most likely to be between 8.3 and 15 GHz most of the time, as for example on 1998 March 16. 
Although sparsely sampled, our IRAM flux densities 
are suggestive of a possible mm flare around 1998 March 18.

The occurrence of this minor flaring event is also supported from the 
hard X-ray (HXR) and soft X-ray (SXR) behaviour of
Cygnus X-3 as recorded by the BATSE (20-100 keV) and ASM (1.5-12 keV)
instruments on board the CGRO and  RXTE satellites, respectively. 
The HXRs are known to be anticorrelated with quiescent
and low level flaring emission in the radio (McCollough et al. 1997).
HXRs are also generally anticorrelated with the SXRs. 
The top panel in Fig. \ref{highe} represents
the one day average from a number of individual ASM dwells. There are
typically 10 dwells of $\sim90$ s each per day, that provide an acceptable sampling  
of the 4.8 h orbital modulation. Therefore, the ASM daily
averages may be considered as representative of the unmodulated level
of emission. Similarly, the bottom panel in this figure contains the
daily averaged HXR data reported by BATSE. We notice here that 
the HRX flux dropped significantly in coincidence with our bright mm
detection, as expected when a low level radio flare occurs.

The 1.25 mm emission on 1998 March 18 was higher than average by a factor of 3. This 
apparent flaring increase
seems to have induced an important evolution in the Cygnus X-3 spectrum. Its main stages
are sketched in Fig. \ref{cx3}. 
We can see here how the high frequency end of the spectrum experiences significative changes.
The 8.3-250 GHz spectral index evolves from $\alpha \simeq -0.6$ during the first three days
(March 14-16) to a nearly flat value during the mm `maximum' (March 18), and back to the original 
negative value after its decay. Unfortunately, the 2 d gap in the Ryle monitoring prevents 
us from interpolating a reliable 15 GHz measurement at the time of the mm `maximum', 
that could provide a better view of this event. 

At low frequencies,  the spectrum remains always optically thick 
with the 2.25-8.3 GHz spectral index being $\alpha \simeq +0.3$.
The slower evolution and consequent overlapping of flaring events at optically thick frequencies
causes the mm flare to be barely evident in the GBI curves. 

\subsection{SS~433}

SS 433 is well known to exhibit highly collimated jets
flowing out at $0.26c$ and precessing every 164 days 
(e.g. Hjellming \& Johnston 1981). 
The accepted interpretation of the observed sub-arcsecond radio structure assumes that, 
during the SS~433 radio outbursts,
twin plasmons or ``bullets", containing relativistic plasma, are ejected from a
central unresolved core into the jet opposite directions (Vermeulen et al. 1987).
The source cm emission is consistent with non-thermal synchrotron radiation 
with optically thin properties. This can be seen clearly in 
the 2 yr GBI radio flux history (see e.g. Fender et al. 1997c). 
At mm wavelengths, a steep optically thin spectrum has been also reported 
(Band \& Gordon 1989; Tsutsumi et al. 1996).

At the time of our observations, it is evident from the left panel in
Fig.~\ref{ss433} that SS~433 was undergoing one of its outburst events at cm wavelengths.
The onset time can be estimated around JD 2450885.5 (1998 March 13), peaking 4 d later
in the GBI data.
At mm wavelengths, our sampling is unfortunately not very good. Nevertheless,  
it is likely that the 250 GHz peak took place between
JD 2450888-889. The mm flux density on these dates 
appears to be higher than in the following days, 
when the outburst decay is already in progress according to GBI.

It is known that SS 433 always exhibits an important optically
thin quiescent emission of non-thermal synchrotron origin. According to
Seaquist et al. (1982), it can be estimated on average as
$S_{\nu}^{\rm quies} = 1.23$~Jy~$(\nu/$GHz$)^{-0.6}$. From the
GBI data during the $\sim20$ d long period of quiescence, prior to 
the outburst event, we derive the
similar power law $S_{\nu}^{\rm quies} = 1.27\pm0.03$~Jy~$(\nu/$GHz$)^{-0.67\pm0.02}$.

In order to better appreciate the spectral evolution of SS~433 during the outburst,
we have subtracted this quiescent component from all spectra
in the right panel of Fig.~\ref{ss433} and, hereafter,
only the flaring emission will be considered. 
The three representative epochs shown correspond to the outburst decaying part, on
JD 2450889, 891 and 892 (March 16, 18 and 19, respectively). For each of these epochs,
we plot two estimates of the intrinsic flaring spectrum. They correspond to the
use of the Seaquist et al. (1982) and the contemporaneous GBI quiescent spectra, respectively.
The difference between both estimates is indicative of the uncertainty involved in
the subtraction process. This problem is only important at the IRAM frequency. 
Probably, the most reliable subtraction is the one involving contemporaneous GBI data.

The results obtained are suggestive that the synchrotron flaring spectrum of
SS~433 extends up to the mm regime throughout the full radio outburst.
In addition we also find possible evidence of spectral steepening during the outburst decay,
at least concerning the GBI frequencies. 
The spectral index between 2.25 and 8.3 GHz varied from
$\alpha = -0.5\pm0.1$ (March 16) to $\alpha = -0.9\pm0.1$ (March 19).
This result does not depend very much on which quiescent spectrum is subtracted.  
On the other hand, the evolution of the spectral index between 8.3 and 250 GHz
also seems to indicate a steepening evolution. Unfortunately, the errors in the
subtraction at 250 GHz prevent us from being very certain about it.




\subsection{LSI+61$^{\circ}$303}

This object is a Be REXRB with periodic and strong outburst events in the radio 
every 26.5 d. This recurrence is assumed to reflect the system 
orbital period and was discovered by Taylor \& Gregory (1982).
LSI+61$^{\circ}$303 is thus the only selected stars where the onset of radio outbursts
can be predicted (see e.g. Paredes et al. 1990).

In the millimetre region, two previous LSI+61$^{\circ}$303 observations are available
in the literature. Altenhoff et al. (1994) observed this star at 250 GHz but they 
did not detect it. Considering their sensitivity, this can be possibly understood
because their observation did not coincide with an outburst episode. On the contrary,
Tsutsumi et al. (1996) do detected LSI+61$^{\circ}$303 at 230 GHz at a flux density level of 10 mJy.
This occurred on two consecutive days close to the expected outburst phases, 
when the source was reported to peak at $\sim100$ mJy at cm wavelengths.

In view of these facts, our observing runs were specially
scheduled in coincidence with one of LSI+61$^{\circ}$303 periodic radio outbursts. 
As it can be seen in the Fig.~\ref{lsi} left panel, the expected flaring event 
actually took place at cm wavelengths according to the GBI monitoring. 
Its amplitude and duration were, however, lower than expected. Variations
in the outburst shape from cycle to cycle are in fact not unusual,
as shown by the extended GBI light curves (see e.g. Ray et al. 1997). 
In our IRAM measurements, we managed to detect the source at
1.25 mm on two consecutive days that happen to be in close coincidence 
with the radio outburst peak.
The right panel in Fig.~\ref{lsi} shows that, at this time (1998 March 15), 
the synchrotron spectrum of LSI+61$^{\circ}$303 extended from cm to mm wavelengths 
over two frequency decades.
A least square fit gives $S_{\nu}=(150\pm3)~$mJy~$(\nu /{\rm GHz})^{-0.47\pm0.01}$
as the best power law representing the data.

During the decay, no mm observation of LSI+61$^{\circ}$303 
resulted in a positive detection and only upper limit estimates are available 
(see Table \ref{obsres}). All these upper limits are consistent with the extrapolation of 
the observed 2.25-8.3 GHz spectral indices in the corresponding GBI data ($\alpha \simeq -0.55$). 
For example, at the times of lowest GBI emission, the expected 250 GHz flux density
should be $\sim2.4$ mJy. This value is below our sensitivity.
  

\subsection{Cygnus X-1}

The cm radio emission of this massive REXRB, and classical black hole candidate,
is usually persistent at the $\sim$ 15 mJy level. Its spectral index is usually flat 
and some radio outbursts, reaching up to 30-40 mJy,
have been occasionally witnessed (Hjellming 1973; Mart\'{\i} et al. 1996).
Radio variations at the binary period (5.6 d) and at a longer  
period (150 d) have been reported (Pooley et al. 1999).
The only mm detection so far published is the value of 10$\pm$3 mJy
given by Altenhoff et al. (1994), suggesting that the generally
flat radio spectrum extends into the mm regime. This suspicion has been
recently confirmed by Fender et al. (1999b), who found no evidence for
a high frequency cut-off up to 220 GHz.

During our monitoring, the centimetric radio emission 
of Cygnus X-1 behaved in the expected way, i.e.,
GBI flux densities in the range $\sim5$-20 mJy with a flat spectral
index being observed. In contrast, very little can be said at 250 GHz. 
The source remained below our IRAM  detection limits throughout 
all our observations. 


\subsection{GRS 1915+105}

This is one of the two superluminal jet sources in the Galaxy as
discovered by Mirabel \& Rodr\'{\i}guez (1994). 
The system has been proposed to be a high mass X-ray binary (Mirabel et
al. 1997). Its radio emission displays highly active episodes, lasting several months
during which multiple and strong ($\sim1$ Jy) radio outbursts have been reported.
During one of these events, a 234 GHz flux density of 123 mJy was observed with the IRAM 30 m
telescope (Rodr\'{\i}guez et al. 1995).
As in SS 433, the GRS 1915+105 flaring episodes correspond to the ejection of twin plasmons,
at relativistic speeds, sometimes producing apparent superluminal motion on the sky.
The most recent sequence of superluminal motion in GRS 1915+105 
has been obtained by Fender et al. (1999c) with unprecedented resolution.

We monitored GRS~1915+105 on five different days at 250 GHz. The level of activity
indicated by the GBI and Ryle monitoring was relatively weak, i.e., a few tens of mJy.
Consistent with this state, no positive 250 GHz detection was achieved in a reliable way.

\subsection{Other X-ray binaries}

On 1998 March 17, most of our program sources were not in good elevation conditions 
during the scheduled time. Therefore, we devoted part of our run
to try to detect three additional X-ray binaries at 250 GHz: the stars 
LSI+65$^{\circ}$010, LSI+61$^{\circ}$235 and X Per.

We observed the first star, LSI+65$^{\circ}$010, for a total of 37 subscans
around 18.5h UT. No positive detection was achieved, with our upper limit estimate 
being $<$17 mJy. Previous attempts to detect this system at cm wavelengths have failed
as well. Nelson \& Spencer (1988) used the Jodrell Bank Lovell-MkII interferometer at 5 GHz
to provide upper limits of $<$1.2 mJy and $<$3.7 mJy in 1986 and 1987, respectively. 
Our group has also carried out cm observations of LSI+65$^{\circ}$010 
with the VLA interferometer of NRAO\footnote{The National Radio Astronomy 
Observatory is a facility of the USA National Science Foundation operated under 
cooperative agreement by Associated Universities, Inc.} on two
consecutive dates, 1993 March 9 and 10. The 20 cm wavelength was used with the
array being in its B configuration. A flux density upper limit
of $<$0.34 mJy was obtained after concatenating all the visibility data.


The second star, LSI+61$^{\circ}$235, was observed for a total number of 80 subscans
around 18.8h UT. Again, no positive detection occurred.
The 250 GHz upper limit in this case is $<$11 mJy. At the 20 cm wavelength, 
a $<0.25$ mJy upper limit is also available for LSI+61$^{\circ}$235
from the same VLA runs mentioned above.
Finally, X Per appeared to be below our detection limits as well
when observed with the IRAM antenna at 18.1h UT.
The number of subscans devoted to it was 60 and the resulting
upper limit $<$15 mJy.

\section{Discussion}

\begin{figure*}[htb]
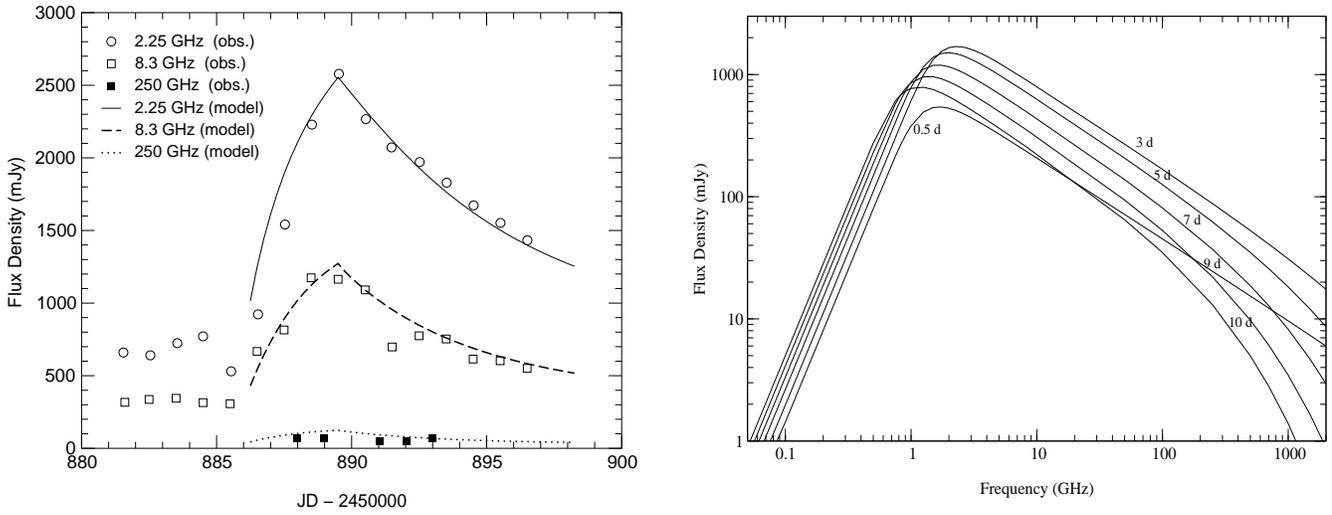

\mbox{}
\vspace{6.5cm}
\includegraphics{aa_hfrad_f5a.ps}
\includegraphics{aa_hfrad_f5b.ps}
\caption[]{{\bf Left.} Theoretical light curves for the SS 433 radio outburst.
They have been computed according to Peracaula (1997). The plasmon parameters adopted are those of Table \ref{param}
and the quiescent spectrum derived from contemporaneus GBI data has been assumed. 
{\bf Right}. Spectral steepening predicted by the model as
result of electron energetic losses. This effect is specially important at high frequencies.}
\label{steep}
\end{figure*}

\subsection{How extended is the synchrotron spectrum in REXRBs?}

We have already mentioned that non-thermal radio emission in REXRBs is attributed to the
ejection of ionized plasmons containing relativistic
electrons. These particles generate synchrotron emission due to the plasmon
magnetic field $B$ acting on them. 
In this scenario, the changes in the radio spectrum are
actually reflecting those occurring in the electron energy distribution. The energy
spectra of the electrons is often well represented by a truncated power law
$N(E)dE=K E^{-p}dE$ ($E_{\rm min} \leq E \leq E_{\rm max}$).
At any time, the synchrotron emission will extend up to a
break frequency $\nu_{\rm break}$ corresponding approximately 
to the critical frequency of electrons
with the highest energy available. In c.g.s. units: 
\begin{equation}
\nu_{\rm break} = 6 \times 10^{18} B E_{\rm max}^2.
\end{equation}

Based on the data of this paper,
the fact that both Cygnus X-3 and SS 433 are persistently detected at 1.25 mm 
implies for these sources that $\nu_{\rm break} > 250$ GHz
most of the time. Consequently, relativistic electrons need to be
accelerated up to energies of $E_{\rm max} > 2.0\times 10^{-4} B^{-1/2}$ erg,
equivalent to gamma factors of $\gamma_{\rm max} > 240 B^{-1/2}$, to account
for the observed high frequency emission.
Estimates of the magnetic field in REXRB plasmons come mainly from 
equipartition arguments, electron age considerations
and theoretical fits to light curves during radio outbursts.
The resulting values are mostly in the range $10^{-3}$-${10^1}$ G 
(see e.g.  Mart\'{\i} et al. 1992; Mart\'{\i} 1993; Mirabel et al. 1998; Ogley et al. 1998).
Adopting $B \sim 10^{-1}$ G as a representative estimate,  
the maximum gamma factor of the electrons is typically expected to be
higher than $\sim10^3$.  These order of magnitude
considerations are also valid for LSI+61$^{\circ}$303 and possibly other REXRBs
when in outburst.

\subsection{Interpreting the spectral steepening in Cygnus X-3 and SS 433}

In both Cygnus X-3 and SS 433, we have witnessed episodes of spectral
steepening during the decaying part of flaring events. As explained 
below, we interpret this behavior in terms of the energy
distribution of the relativistic electrons evolving in time during
a flaring event. Of course we cannot rule out that other effects
may be work, although the proposed explanation appears to be reproduce 
this kind of behavior.

\begin{table}
\caption[]{\label{param} Plasmon physical parameters for the SS 433 radio outburst}
\begin{tabular}{ll}
\hline
Initial radius            &   $3.0 \times 10^{14}$ cm \\
Expansion velocity        &   1650 km s$^{-1}$        \\
Initial magnetic field    &   1.8 G     \\
Injection time interval   &  3.5 d                     \\ 
Electron power law index  &  2.3                      \\
Electron injection rate   &  $1.3 \times 10^{-15}$ $M_{\odot}$ d$^{-1}$ \\
Energy limits             &  $\gamma=1$ and $6 \times 10^4$     \\
\hline
\end{tabular}
\end{table}

Our interpretation is as follows.
Since the electrons are undergoing energetic losses of different kind (expansion losses,
synchrotron radiation, inverse Compton scattering, etc.), the distribution $N(E)dE$
will be shifted towards lower energies. The values of $E_{\rm max}$, and consequently
that of $\nu_{\rm break}$ will be decreasing functions of time. 
A progressive steepening of the radio spectrum is thus expected as the outburst
evolves and decays. Theoretical modeling of plasmon evolution, within spherical
and jet geometries, has been addressed among others by Paredes et al. (1991),
Mart\'{\i} et al. (1992) and Peracaula (1997). When these models are used to
compute multi-epoch spectra, we are able to reproduce a clear steepening behavior  
in strong resemblance to what is observed.

The plots in Fig. \ref{steep} are intended to be
an illustrative example of what we have described in a qualitative way.
They correspond to a modelling attempt of the SS 433 radio outburst
previously discussed, i.e., the best sampled event that is well suitable for this purpose.
The best fit plasmon physical parameters found
are listed in Table \ref{param}.  The Paredes et al. (1991) formulation 
has been used in the calculations as improved by Peracaula et al. (1997).
Two plasmons ejected into opposite directions at a velocity of $0.26c$ are
considered. The ejection center coincides
with the compact companion of the binary system with an adopted orbital
separation of $1.4\times 10^{15}$ cm. The assumed luminosity of the
optical companion is $4 \times 10^{39}$ erg s$^{-1}$. It is this radiation
field which causes most of the steepening seen in Fig. \ref{steep}
through strong inverse Compton losses.

\section{Conclusions}

\begin{enumerate}

\item We have conducted a series of millimetre and centimetre observations
of a sample of REXRBs during a week long interval. 
Our target list included both Cygnus X-3 and
SS 433, that were persistently detected at 250 GHz (1.25 mm) throughout 
the whole run. The REXRB LSI+61$^{\circ}$303 was also detected at 250 GHz 
near the peak of one of its periodic radio outbursts. 
Several 250 GHz upper limits for other REXRBs are also reported.

\item For the detected sources, our results are in agreement with
the synchrotron spectrum in REXRBs extending commonly 
up to millimetre wavelengths and possibly beyond. This observed fact
reinforces the idea that these systems are able to accelerate
relativistic electrons to very high energies, 
at least $\gamma \geq 10^3$.
 
\item The high frequency radio spectrum of Cygnus X-3 and SS 433 was also
observed to steepen noticeably during the decay of flaring events. This
behavior is interpreted in terms 
of energetic losses of the synchrotron emitting electrons.

\end{enumerate}

\begin{acknowledgements}
JMP and JM acknowledge partial support by DGICYT (PB97-0903). JM is in addition 
supported by Junta de Andaluc\'{\i}a (Spain), and wishes to thank as well
the hospitality and support of the Service d'Astrophysique (CEA/Saclay, France)
during the early stages of this work.

\end{acknowledgements}

\end{document}